\documentclass[10pt]{article}
\usepackage[T1]{fontenc}
\usepackage[latin1]{inputenc}
\usepackage{amsmath}
\usepackage{graphics}
\usepackage{setspace}

\makeatletter

\newcommand{\LyX}{L\kern-.1667em\lower.25em\hbox{Y}\kern-.125emX\spacefactor1000}
\newenvironment{LyXParagraphIndent}[1]%
{
  \begin{list}{}{%
    \setlength\topsep{0pt}%
    \addtolength{\leftmargin}{#1}
    \setlength\parsep{0pt plus 1pt}%
  }
  \item[]
}
{\end{list}}

\renewcommand\theparagraph    {\@arabic\c@paragraph } 
\makeatother
\begin{document}
\newenvironment{Proof}{\begin{quote}\footnotesize {\bf Proof}\hskip 
0.3cm}{\end{quote}\normalsize } 

\input{epsf.tex} 

~

~

~

\noindent \textbf{\Large On Fine's resolution of the EPR-Bell problem}{\Large \par}

~

\begin{LyXParagraphIndent}{1cm}
\noindent {\bf L\'{a}szl\'{o} E. Szab\'{o}}\footnote{
Theoretical Physics Research Group of HAS and Department of History and Philosophy of Science, E\"{o}tv\"{o}s University, Budapest
}

\end{LyXParagraphIndent}

\begin{LyXParagraphIndent}{1cm}
\begin{quotation}
\noindent ~

\noindent ~
\end{quotation}\end{LyXParagraphIndent}

\begin{LyXParagraphIndent}{1cm}
\noindent \( \overline{\, \, \, \, \, \, \, \, \, \, \, \, \, \, \, \, \, \, \, \, \, \, \, \, \, \, \, \, \, \, \, \, \, \, \, \, \, \, \, \, \, \, \, \, \, \, \, \, \, \, \, \, \, \, \, \, \, \, \, \, \, \, \, \, \, \, \, \, \, \, \, \, \, \, \, \, \, \, \, \, \, \, \, \, \, \, \, \, \, \, \, \, \, \, \, \, \, \, \, \, \, \, \, \, \, \, \, \, \, \, \, \, \, \, \, \, \, \, \, \, \, \, \, \, \, \, \, \, \, \, \, \, \, \, \, \, \, \, \, \, \, \, \, \, \, \, \, \, \, \, \, \, \, \, \, \, \, \, \, \, \, \, \, \, \, \, \, \, \, \, \, \, \, \, \, \, } \)\\
\textit{\small In the real spin-correlation experiments the detection/emission
inefficiency is usually ascribed to independent random detection errors, and treated by the 
``enhancement hypothesis''. In Fine's
``prism model'' the detection inefficiency is an effect not only of the random
errors in the analyzer + detector equipment, but is also the manifestation of
a pre-settled (hidden) property of the particles.} \\ \( \underline{\, \, \, \, \, \, \, \, \, \,
\, \, \, \, \, \, \, \, \, \, \, \, \, \, \, \, \, \, \, \, \, \, \, \, \, \,
\, \, \, \, \, \, \, \, \, \, \, \, \, \, \, \, \, \, \, \, \, \, \, \, \, \,
\, \, \, \, \, \, \, \, \, \, \, \, \, \, \, \, \, \, \, \, \, \, \, \, \, \,
\, \, \, \, \, \, \, \, \, \, \, \, \, \, \, \, \, \, \, \, \, \, \, \, \, \,
\, \, \, \, \, \, \, \, \, \, \, \, \, \, \, \, \, \, \, \, \, \, \, \, \, \,
\, \, \, \, \, \, \, \, \, \, \, \, \, \, \, \, \, \, \, \, \, \, \, \, \, \,
\, \, \, \, \, \, \, \, \, \, } \)

\end{LyXParagraphIndent}
\setlength{\unitlength}{1cm} ~\begin{picture}(5,1)(2,-10)
\put(0,0){E\"otv\"os-HPS 99-24} \end{picture}

\section{\textsc{Introduction}}

The aim of this paper is to make an introduction to Fine's interpretation of
quantum mechanics and to show how it can solve the EPR--Bell problem. In the
real spin correlation experiments the measured quantum probabilities are identified
with relative frequencies taken on a selected sub-ensemble of the emitted particle
pairs: only those particles are taken into account which are coincidentally
detected in the two wings. The detection/emission inefficiency is usually ascribed
to random detection errors occurring independently in the two wings, and treated
by the ``enhancement hypothesis'', which assumes that the relative frequencies
measured on the randomly selected sub-ensemble are equal to the ones taken on
the whole statistical ensemble of emitted particle pairs. 

Fine's \emph{prism model}\( ^{(1)} \) is a local hidden variable theory, in
which the detection inefficiency is an effect not only of the random errors
in the analyzer + detector equipment, but is also the manifestation of a predetermined
hidden property of the particles.\footnote{
This conception of hidden variable goes back to Einstein (Ref. 4, Chapter 4).
} I present one of Fine's prism models for the EPR experiment and compare it
with the recent experimental results.\( ^{(2)} \) As we shall see, it works
well in case of the \( 2\times 2 \) spin-correlation experiments. There appeared,
however, a theoretical demand to embed the \( 2\times 2 \) prism models into
a large \( n\times n \) prism model reproducing all potential \( 2\times 2 \)
sub-experiments. This demand was motivated by the idea that the real physical
process does not know which directions are chosen in an experiment. On the other
hand, it seemed that in the known prism models of the \( n\times n \) spin-correlation
experiment the efficiencies tended to zero, if \( n\rightarrow \infty~ \), which
contradicts what we expect of actual experiments.\( ^{(3)} \) This problem
is investigated in the last part of the paper.

\section{Real EPR experiments}

\begin{figure}
{\centering \resizebox*{0.8\columnwidth}{!}{\includegraphics{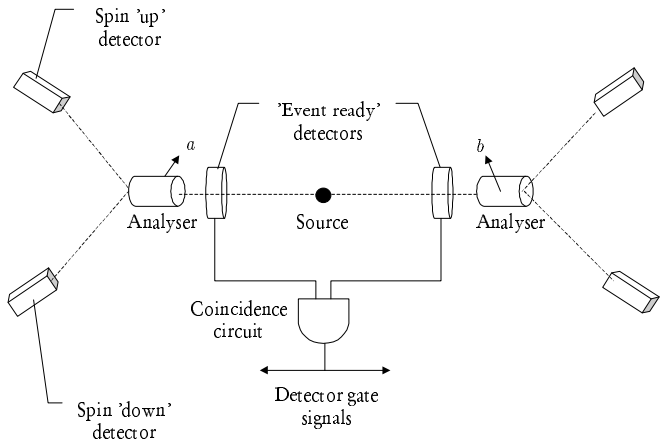}} \par}

\caption{\emph{Apparatus configuration used for Bell's 1971 proof. `Event-ready' detectors
signal both arms that a pair of particles has been emitted\label{fig_Bell}}}
\end{figure}

Figure~\ref{fig_Bell} shows an Aspect-type spin-correlation experiment. The
analyzers can be set in orientation \( a \) or \( a' \) on the left hand side,
and \( b \) or \( b' \) on the right. Denote \( A \), \( A' \), \( B \)
and \( B' \) the corresponding ``spin-up'' detection-events. \( p(A|a) \),
for instance, denotes the conditional probability of \( A \), given that the
measurement set-up in the left wing is \( a \). 

Now, from the assumption that there exists a hidden variable \( \lambda  \)
satisfying the screening off condition,
\[
p\left( A\wedge B|a\wedge b\wedge \lambda \right) =p\left( A|a\wedge \lambda \right) p\left( B|b\wedge \lambda \right) \]

\noindent one can derive\( ^{(4)} \) the well known Clauser-Horne inequality:
\begin{equation}
\label{eq_CH}
-1\leq \underbrace{\left\{ \begin{array}{c}
p\left( A\wedge B|a\wedge b\right) +p\left( A\wedge B'|a\wedge b'\right) -p\left( A'\wedge B|a'\wedge b\right) \\
+p\left( A'\wedge B'|a'\wedge b'\right) -p\left( A'|a'\right) -p\left( B|b\right) 
\end{array}\right\} }_{CH}\leq 0
\end{equation}
 According to the standard views, inequality (\ref{eq_CH}) is violated in the
real spin-correlation experiments, hence, the argument goes, any local hidden
variable theory for the EPR experiment is excluded. For example, in case of
spin-\( \frac{1}{2} \) particles, if the directions \( a \), \( a' \), \( b \),
\( b' \) are coplanar with \( \angle (a,b)=\angle (a,b')=\angle (a',b')=120^{\circ } \)
and \( \angle (a',b)=0 \), then the Clauser-Horne expression in (\ref{eq_CH})
is 
\begin{equation}
\label{eq_CH_szamok}
CH=\frac{3}{8}+\frac{3}{8}-0+\frac{3}{8}-\frac{1}{2}-\frac{1}{2}=\frac{3}{8}
\end{equation}

\begin{figure}
{\centering \resizebox*{0.8\columnwidth}{!}{\includegraphics{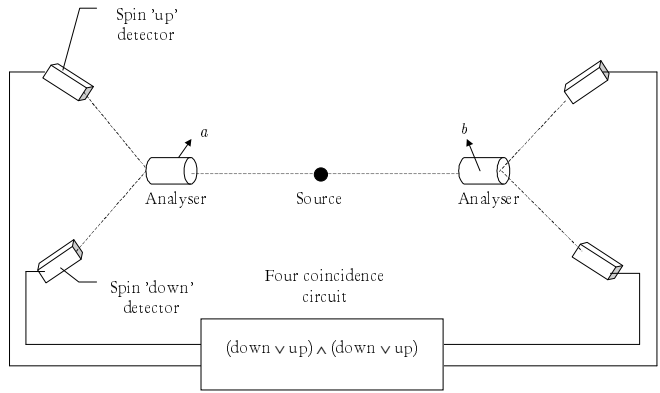}} \par}
\caption{\emph{In the real experiments, instead of the event-ready detectors, a four-coincidence circuit detects the `emitted particle-pairs'\label{fig_valosagos_EPR}}}
\end{figure}

There is, however a serious loophole in the real experiments. Compare the original
apparatus configuration used for Bell's 1971 proof with the one used in the
real Aspect experiment (Fig.~\ref{fig_Bell} and \ref{fig_valosagos_EPR}). The
original configuration contains two `event-ready' detectors, which signal both
arms that a pair of particles has been emitted. So, the statistics are taken
on the ensemble of particle pairs emitted by the source. In the real experiments,
however, instead of the event-ready detectors, a four-coincidence circuit detects
the `emitted particle-pairs'. This method yields to a \emph{selected} statistical
ensemble: only those pairs are taken into account, which coincidentally fire
one of the left and one of the right detectors. Denote \( \left[ A\right]  \)
the event that there is any detection in the left wing with analyzer set-up
\( a \), that is, either the ``up'' detector or the ``down'' detector fires.
Similarly, \( \left[ A\right] \wedge \left[ B\right]  \) denotes the corresponding
double detection. So, what we actually observe is the violation of the following
inequality:

\begin{equation}
\label{eq_CHobserved}
-1\leq \left\{ \begin{array}{c}
p\left( A\wedge B|a\wedge b\wedge \left[ A\right] \wedge \left[ B\right] \right) +p\left( A\wedge B'|a\wedge b'\wedge \left[ A\right] \wedge \left[ B'\right] \right) \\
-p\left( A'\wedge B|a'\wedge b\wedge \left[ A'\right] \wedge \left[ B\right] \right) +p\left( A'\wedge B'|a'\wedge b'\wedge \left[ A'\right] \wedge \left[ B'\right] \right) \\
-p\left( A'|a'\wedge \left[ A'\right] \right) -p\left( B|b\wedge \left[ B\right] \right) 
\end{array}\right\} \leq 0
\end{equation}
 If the selection procedure were \emph{completely random} then the observed
relative frequencies on the selected ensemble would be equal to the ones taken
on the original  ensemble, that is,
\begin{eqnarray*}
p\left( A\wedge B|a\wedge b\wedge \left[ A\right] \wedge \left[ B\right] \right)  & = & p\left( A\wedge B|a\wedge b\right) \\
p\left( A\wedge B'|a\wedge b'\wedge \left[ A\right] \wedge \left[ B'\right] \right)  & = & p\left( A\wedge B'|a\wedge b'\right) \\
 & \textrm{etc}. & 
\end{eqnarray*}
\noindent (\emph{enhancement} \emph{hypothesis}) and the violation of inequality
(\ref{eq_CHobserved}) would imply the violation of (\ref{eq_CH}), in accordance with Bell's point of view:
\begin{quote}
\noindent ... it is hard for me to believe that quantum mechanics works so nicely
for inefficient practical set-ups and is yet going to fail badly when sufficient
refinements are made. (Ref. 5, p. 154)
\end{quote}
This is indeed the case if non-detections are caused by independent random errors in
the detector+analyser equipment.

\section{Fine's interpretation}
Arthur Fine approaches the detection inefficiency problem in a different way: 
\begin{quote}
\noindent ... the efficiency problem ought not to be dismissed as merely one
of biased statistics and conspiracies, for the issue it raises is fundamental.
Can a hidden variable theory of the very type being tested explain the statistical
distributions, inefficiencies and all, actually found in the experiments? If
so then we would have a model (or theory) of the experiment that explains why
the samples counted yield the particular statistics that they do. (Ref. 6, p.
465) 
\end{quote}
This conception of hidden variable was first realized in Fine's \emph{prism
models}\( ^{(1)} \) for the \( 2\times 2 \) spin-correlation experiment. Prism
model is a local, deterministic hidden variable theory, in which the hidden
variables predetermine not only the outcomes of the corresponding measurements,
but also predetermine whether or not an emitted particle arrives to the detector
and becomes detected. In other words, the measured observables can take on a
new ``value'' corresponding to an inherent ``no show'' or defectiveness. 
\begin{figure}[t] 
  	\begin{center}\leavevmode 
	\hskip1cm\epsfxsize=8cm 
	\epsfbox{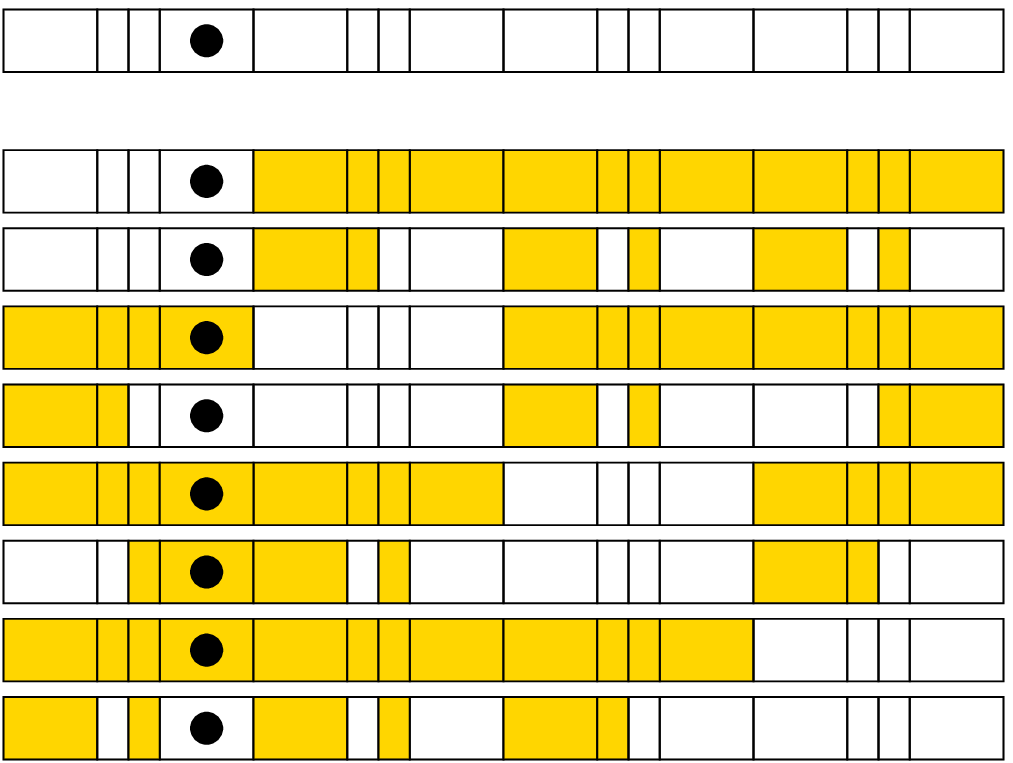} 
  	\end{center} 
\setlength{\unitlength}{0.1cm} 
\begin{picture}(0,0)(0,0) 
\put(68,71){\line(-1,-1){5}}
\put(70,73){paramter space $\Lambda\ni\lambda$} 
\put(35,59){ $\lambda = \lambda_{example}$}
\put(42,71){\line(-1,-1){5}}
\put(44,73){$\frac{1}{32}$} 
\put(23,71){\line(1,-1){5}}
\put(17,73){$\frac{3}{32}$} 
\put(15,53){ $[A]$}
\put(15,47){ $A$}
\put(15,40){ $[A']$}
\put(15,34){ $A'$}
\put(15,28.5){ $[B]$}
\put(15,21.6){ $B$}
\put(15,15.5){ $[B']$}
\put(15,9){ $B'$}
\end{picture}
\caption{\em Parameter $\lambda\in\Lambda$ completely predetermines all future
events for all possible combinations of the freely chosen measurements. Each EPR event is represented as a (shaded) subset of \( \Lambda\). }  
\label{fig_prism}
\end{figure}

As an example, consider a prism model reproducing the quantum mechanical probabilities
in (\ref{eq_CH_szamok}). Figure~\ref{fig_prism} shows a parameter space \( \Lambda \ni \lambda \) consisting of disjoint blocks of measure \( \frac{3}{32} \) and \( \frac{1}{32} \) respectively. A point of \( \Lambda \) (a value of the parameter \( \lambda \)) predetermines all events in question. Therefore, each EPR event can be represented as a subset of \( \Lambda\). For instance, assume that \( \lambda =\lambda _{\mathrm{example}} \). Then, an \( a \)-measurement on the left particle produces neither event ``up'' nor event ``down'', while if an \( a' \)-measurement is performed then the outcome is ``down''. In the right wing, if we perform a \( b \)-measurement then the outcome is ``up'', and if the \( b' \)-measurement is performed, the outcome is ``down''. Consequently, in case, for example, we perform an \( a \)-measurement on the left particle and a \( b \)-measurement on the right one, then there is no coincidence registered, and the particle pair in question does not appear in the statistics of the measurement. On the contrary, if we perform an \( a' \)-measurement on the left particle and a \( b \)-measurement on the right one, then there is a coincidence registered and the counter of the total number of events as well as the \( B \)-counter count. Thus, the hidden parameter governs the whole process in such a way that the observed relative frequencies reproduce the probabilities measured in the experiment:
\begin{eqnarray*}
\frac{\mu \left( A\right) }{\mu \left( [A]\right) } & = & \frac{\mu \left(
A'\right) }{\mu \left(  [A']\right) }=\frac{\mu \left( B\right)
}{\mu \left( [B] \right) }=\frac{\mu \left( B'\right) }{\mu \left( 
[B']\right) }=\frac{\frac{12}{32}}{\frac{24}{32}}=\frac{1}{2}\\  \frac{\mu
\left( A\cap B\right) }{\mu \left( [A]\cap [B]\right) } & = &
\frac{\mu \left( A\cap  B'\right) }{\mu \left( [A]\cap [B']\right)
}=\frac{\mu \left( A'\cap B'\right) }{\mu \left(  [A']\cap
[B']\right) }=\frac{\frac{6}{32}}{\frac{16}{32}}=\frac{3}{8}\\  \frac{\mu
\left( A'\cap B\right) }{\mu \left( [A']\cap [B]\right) } & = &
\frac{0}{\frac{16}{32}}=0  \end{eqnarray*}

\section{Compatibility with the actual EPR experiments}
\label{section_recent_experiments_1}As we have seen, the basic idea of
the Einstein--Fine interpretation is that some elements of the statistical ensemble
of identically prepared quantum systems (characterized by a quantum state \( W \))
do not produce outcome at all when one perform the measurement of a quantum
observable \( A \). Such systems are called \( A \)\emph{-defective} in Fine's
terminology. In connection with this basic feature of the model, one can investigate
some important characteristics of the above Einstein--Fine model of the EPR
experiment, and compare them with the similar characteristics of the actual
EPR experiments: 
\begin{eqnarray*}
R^{A} & = & \frac{\textrm{number of non}-A-\textrm{defective systems}}{\textrm{total number of systems}}\\
R^{A'} & = & \frac{\textrm{number of non}-A'-\textrm{defective systems}}{\textrm{total number of systems}}\\
R^{B} & = & \frac{\textrm{number of non}-B-\textrm{defective systems}}{\textrm{total number of systems}}\\
R^{B'} & = & \frac{\textrm{number of non}-B'-\textrm{defective systems}}{\textrm{total number of systems}}\\
R^{AB} & = & \frac{\textrm{number of non}-A-\textrm{defective }\&\textrm{ non}-B-\textrm{defective systems}}{\textrm{total number of systems}}\\
R^{AB'} & = & \frac{\textrm{number of non}-A-\textrm{defective }\&\textrm{ non}-B'-\textrm{defective systems}}{\textrm{total number of systems}}\\
R^{A'B} & = & \frac{\textrm{number of non}-A'-\textrm{defective }\&\textrm{ non}-B-\textrm{defective systems}}{\textrm{total number of systems}}\\
R^{A'B'} & = & \frac{\textrm{number of non}-A'-\textrm{defective }\&\textrm{ non}-B'-\textrm{defective systems}}{\textrm{total number of systems}}
\end{eqnarray*}
In case of the above example:
\begin{equation}
\label{eq_maximal_prism_efficiency}
\begin{array}{rcl}
R^{A}=R^{A'}=R^{B}=R^{B'} & = & 75\%\\
R^{AB}=R^{AB'}=R^{A'B}=R^{A'B'} & = & 50\%
\end{array}
\end{equation}
If any of the similar rates in a real experiment were higher than the corresponding
one in (\ref{eq_maximal_prism_efficiency}), Fine's interpretation
would be \emph{experimentally}  refuted. 

There are principal
obstacles to an event ready detection, therefore we cannot have a precise information
about the total number of systems. In one of the best experiments of the last
years$^{(2)}$ the \emph{estimated}  rates are the following\footnote{
I would like to thank G. Weihs and A. Zeilinger for the private communications
about many interesting details of the experiment.}:
\begin{equation}
\label{eq_experimental_rates}
\begin{array}{rcl}
R^{A}=R^{A'}=R^{B}=R^{B'} & = & 5\%\\
R^{AB}=R^{AB'}=R^{A'B}=R^{A'B'} & = & 0.25\%
\end{array}
\end{equation}
 So the prism model is in accordance with this experiment.
 
It can be (and probably is) the case that this very low detection/emission rate is mostly  caused by the external random detection errors, different from the prism mechanism. In order to separate these two sources of inefficiency, consider
a new characteristic of the model:
\[
r_{A}^{AB}=\frac{\textrm{number of non}-A-\textrm{defective }\&\textrm{ non}-B-\textrm{defective systems}}{\textrm{number of non}-A-\textrm{defective systems}}\]
\[
r_{A}^{AB'}=\frac{\textrm{number of non}-A-\textrm{defective }\&\textrm{ non}-B'-\textrm{defective systems}}{\textrm{number of non}-A-\textrm{defective systems}}\]
\begin{equation}
\label{eq_rates_2}
\vdots 
\end{equation}
\[
r_{B'}^{A'B'}=\frac{\textrm{number of non}-A'-\textrm{defective }\&\textrm{ non}-B'-\textrm{defective systems}}{\textrm{number of non}-B'-\textrm{defective systems}}\]
In our prism model:
\[
r_{A}^{AB}=r_{A}^{AB'}=r_{A'}^{A'B}=r_{A'}^{A'B'}=r_{B}^{AB}=r_{B'}^{AB'}=r_{B}^{A'B}=r_{B'}^{A'B'}=66,66\%\]
The experiment by Weihs \emph{et al}.$^{(2)}$  had a particular new feature: In
the two wings independent data registration was performed by each observer having
his own atomic clock, synchronized only once before each experiment cycle. A
time tag was stored for each detected photon in two separate computers at the
observer stations and the stored data were analyzed for coincidences long after
measurements were finished. Due to this method of data registration, it was
possible to count the rates in (\ref{eq_rates_2}). Again, if any of these rates
were higher than 66,66\%, Fine's interpretation wouldn't be tenable.
However the experimental values were only around 5\%.

\section{The prism model of an \protect\( n\times n\protect \) spin correlation experiment} \label{section_nxn_problem}
Let us turn now to a serious objection
to Fine's approach.
In the EPR experiment we consider only \( 2\times 2 \) different possible directions
(\( \overrightarrow{a},\overrightarrow{a}',\overrightarrow{b},\overrightarrow{b}' \)).
If nature works according to Fine's prism model then there must exist, in principle,
a larger \( n\times n \) prism model reproducing all potential \( 2\times 2 \)
sub-experiments. It is because nature does not know about how the experiment
is designed. So, in the final analysis, there is no such a thing as \( 2\times 2 \) prism
model of the \( 2\times 2 \) spin-correlation experiment. If we want to describe
a \( 2\times 2 \) spin-correlation experiment with a prism model, then there
must exist a large \( n\times n \) (if not \( \infty \times \infty ) \) prism
model behind it.

The general schema of the prism model of a spin-correlation experiment is the
following. In both wings one considers \( n \) different possible events:
\begin{equation}
\label{eq_nxn_esemenyek}
\overbrace{\underbrace{A_{1},A_{2}}_{\left[ A_{1}\right] =\left[ A_{2}\right] },\underbrace{A_{3},A_{4}}_{\left[ A_{3}\right] =\left[ A_{4}\right] },\ldots \underbrace{A_{n-1},A_{n}}_{\left[ A_{n-1}\right] =\left[ A_{n}\right] }}^{\mathrm{left}}\, \, \, \, \, \overbrace{\underbrace{B_{1},B_{2}}_{\left[ B_{1}\right] =\left[ B_{2}\right] },\underbrace{B_{3},B_{4}}_{\left[ B_{3}\right] =\left[ B_{4}\right] },\ldots \underbrace{B_{n-1},B_{n}}_{\left[ B_{n-1}\right] =\left[ B_{n}\right] }}^{\mathrm{right}}
\end{equation}

\noindent \( A_{1} \) denotes the event that the left particle has spin ``up''
along direction \( \overrightarrow{a}_{1} \). \( A_{2} \) denotes the event
that the left particle has spin ``down'' along direction \( \overrightarrow{a}_{1} \).
Similarly, \( A_{3} \) denotes the event that the left particle has spin ``up''
along direction \( \overrightarrow{a}_{3} \) and \( A_{4} \) denotes the event
that the left particle has spin ``down'' along direction \( \overrightarrow{a}_{3} \),
etc. (We will also use the following notation: \( \overrightarrow{a}_{2}=\overrightarrow{a}_{1},\overrightarrow{a}_{4}=\overrightarrow{a}_{3},\dots \overrightarrow{a}_{2k}=\overrightarrow{a}_{2k-1},\dots \overrightarrow{b}_{2k}=\overrightarrow{b}_{2k-1} \).)
There are \( \frac{n}{2} \) different directions on both sides. We also assume
the following logical relationships:
\begin{equation}
\label{eq_nxn_tulajdonsagok}
\begin{array}{rcllrcl}
\left[ A_{1}\right] =\left[ A_{2}\right]  & = & A_{1}\vee A_{2} &  & \left[ B_{1}\right] =\left[ B_{2}\right]  & = & B_{1}\vee B_{2}\\
\left[ A_{3}\right] =\left[ A_{4}\right]  & = & A_{3}\vee A_{4} &  & \left[ B_{3}\right] =\left[ B_{4}\right]  & = & B_{3}\vee B_{4}\\
 & \vdots  &  &  &  & \vdots  & \\
\left[ A_{n-1}\right] =\left[ A_{n}\right]  & = & A_{n-1}\vee A_{n} &  & \left[ B_{n-1}\right] =\left[ B_{n}\right]  & = & B_{n-1}\vee B_{n}\\
A_{1}\wedge A_{2} & = & 0 &  & B_{1}\wedge B_{2} & = & 0\\
A_{3}\wedge A_{4} & = & 0 &  & B_{3}\wedge B_{4} & = & 0\\
 & \vdots  &  &  &  & \vdots  & \\
A_{n-1}\wedge A_{n} & = & 0 &  & B_{n-1}\wedge B_{n} & = & 0
\end{array}
\end{equation}
 that is, \( \left[ A_{1}\right]  \) (which is equal to \( \left[ A_{2}\right]  \))
denotes the event that the left particle is predetermined to produce any outcome
if the \( \overrightarrow{a}_{1} \) direction is measured. The quantum probabilities
are reproduced in the following way:
\begin{eqnarray}
tr\left( WP_{A_{i}}\right) =q_{i} & = & \frac{p\left( A_{i}\right) }{p\left( \left[ A_{i}\right] \right) }\nonumber \\
tr\left( WP_{B_{i}}\right) =q'_{i} & = & \frac{p\left( B_{i}\right) }{p\left( \left[ B_{i}\right] \right) }\label{eq_nxn_quantum_prob} \\
tr\left( WP_{A_{i}}P_{B_{j}}\right) =q_{ij} & = & \frac{p\left( A_{i}\wedge B_{j}\right) }{p\left( \left[ A_{i}\right] \wedge \left[ B_{j}\right] \right) }\nonumber 
\end{eqnarray}
 The quantum probabilities \( q_{1},q_{2},\ldots q'_{1},q'_{2},\ldots q_{ij},\ldots  \)
are the only fix numbers in the model. 

The experimental setup shows the following simple and natural symmetries:
\begin{itemize}
\item [(S1)]None of the left and right wings is privileged. 
\item [(S2)]There is no privileged direction among the possible polariser positions. 
\end{itemize}
Consequently, all physically plausible prism model have to satisfy these symmetry conditions, which imply the following two constraints:
\begin{equation}
\label{eq_Fine_symmetry_1}
p\left( \left[ A_{i}\right] \right) =\omega \quad \quad \textrm{for all }1\leq i\leq n
\end{equation}
 \noindent where \( \omega  \) is some uniform efficiency for all directions
on both sides, and
\begin{equation}
\label{eq_Fine_symmetry_2}
p\left( \left[ A_{i}\right] \wedge \left[ B_{j}\right] \right) = \sigma \left( \angle \left( \overrightarrow{a}_{i},\overrightarrow{b}_{j}\right) \right) \quad \quad \textrm{for all }1\leq i,j\leq n
\end{equation}
\noindent where \( 0\leq \sigma \left( \angle \left( \overrightarrow{a}_{i},\overrightarrow{b}_{j}\right) \right) \leq \omega \)
is an arbitrary function of the angel \( \angle \left( \overrightarrow{a}_{i},\overrightarrow{b}_{j}\right)  \).

Thus, the
prism-model resolution of the EPR-Bell problem requires the existence of \(
n\times n \) prism models of the above type. On the other hand, this
requirement appears to be  a serious objection to Fine's program. The reason
is that in all the known  $n\times n$ prism models  the efficiency $\omega $
tends to zero if $n\rightarrow \infty$, which contradicts the recent
experimental results.$^{(7-8)} $  Moreover,  Fine  has shown$^{(3)} $ that
this is true for the class of \( n\times n \) prism models satisfying certain
symmetry conditions called \emph{Exchangeability, Indifference} and
\emph{Strong Symmetry}. They are complex conditions,  too complex to briefly
recall the definitions. Although they do not express some natural and obvious
symmetries of the experimental setup, they are instanced in all the  known
prism models. 

If all physically plausible prism models had to satisfy the Exchangeability, Indifference and 
Strong Symmetry conditions, the problem of zero  efficiency would mean a
serious objection to Fine's interpretation.  Fortunately,  this is not the
case. In my http://arXiv.org/abs/quant-ph/0012042 I shown the existence of \(
\infty\times \infty \) prism models satisfying the symmetry conditions (S1)
and (S2), whereas \emph{the efficiencies are reasonably high.}

\section*{Acknowledgments}

The author wish to thank Professor A. Fine for his stimulating comments and
suggestions.  The research was supported by the OTKA Foundation, No. T015606 and T032771.

\section*{\textsc{References}}

\begin{enumerate}
\item A. Fine, ``Some local models for correlation experiments'', \emph{Synthese},
\textbf{50}, 279 (1982)
\item G. Weihs, T. Jennewin, C. Simon, H. Weinfurter, and A. Zeilinger, ``Violation
of Bell's Inequality under Strict Einstein Locality Conditions'', \emph{Phys. Rev. Lett.},
\textbf{81}, 5039 (1998)
\item A. Fine, ``Inequalities for Nonideal Correlation Experiments'', \emph{Foundations of Physics},
\textbf{21}, 365 (1991)
\item B. C. van Frassen, ``The Charybdis of Realism: Epistemological Implications
of Bell's Inequality'', in \emph{Philosophical Consequences of Quantum Theory
-- Reflections on Bell's Theorem}, J. T. Cushing and E. McMullin (eds.), University
of Notre Dame Press, Notre Dame, (1989)
\item J. S. Bell, \emph{Speakable and unspeakable in quantum mechanics}, Cambridge
University Press, Cambridge, (1987)
\item A. Fine, ``Correlations and Efficiency: Testing Bell Inequalities'', \emph{Foundations
of Physics}, \textbf{19}, 453 (1989)
\item  W. D. Sharp  and N. Shank,  , ``Fine's prism models for quantum
correlation statistics'', \emph{Philosophy of Science}, \textbf{52}, 538 (1985)
\item T. Maudlin, \emph{Quantum Non-Locality and Relativity}  -- \emph{Metaphysical Intimations of Modern Physics}, Aristotelian Society Series, Vol. 13, Blackwell, Oxford, (1994)

\end{enumerate}
\end{document}